\documentclass[%
 reprint,
 amsmath,amssymb,
 aps, prl,
]{revtex4-2}

\usepackage[utf8]{inputenc}
\usepackage{verbatim}
\usepackage{graphicx}
\usepackage{dcolumn}
\usepackage{bm}
\usepackage{xspace}
\usepackage{hyperref}

\newcommand{\perkeo}{\textsc{Perkeo~III}\xspace}

\begin{document}

\preprint{APS/123-QED}

\title{Gaussian Processes and Bayesian Optimization for High Precision Experiments}%

\author{M. Lamparth}
  \email{max.lamparth@tum.de}
\author{M. Bestehorn}
\author{B. Märkisch}
\affiliation{
 Physik-Department ENE, Technische Universität München, Germany
}

\date{May 27, 2022}

\begin{abstract}
\noindent
High-precision measurements require optimal setups and analysis tools to achieve continuous improvements.
Systematic corrections need to be modeled with high accuracy and known uncertainty to reconstruct underlying physical phenomena.
To this end, we present Gaussian processes for modeling experiments and usage with Bayesian optimization, on the example of an electron energy detector, achieving optimal performance.
We demonstrate the method’s strengths and outline stochastic variational Gaussian processes for physics applications with large data sets, enabling new solutions for current problems.
\end{abstract}

\maketitle

\begin{figure}[t]
    \includegraphics[width=0.4\textwidth]{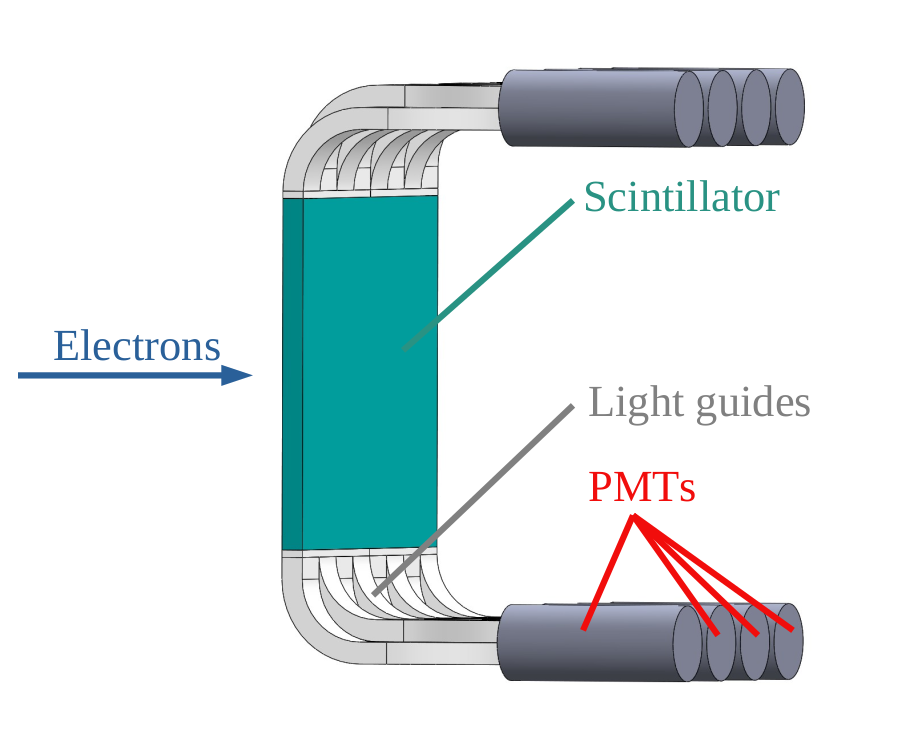}
    \caption{\label{fig:det_schematic} Electron energy detector schematic. The spatial response measurement grid points are in front of the detector and parallel to the scintillator surface.}
\end{figure}

Particle physics experiments have many tunable parameters and are designed for specific purposes, leading to complex designs.
We need to model an experiment at all stages for precise measurement results.
We must achieve optimal experiment design before its construction, run it efficiently and understand underlying phenomena during data analysis.
As a specific example, we will optimize the analysis of an electron energy detector by improving its calibration by modeling different systematic effects and their uncertainty.
To this end, we use a generalization of multivariate normal distributions, called Gaussian processes \cite{williams1995gaussian, GpRasWil}.
They generalize the probability density functions of random vector variables to functions by using Bayesian inference and conditioning on already observed data.
Gaussian processes are constructed from an observed data set itself and grow in complexity with the data set, making them non-parametric and robust to overfitting \cite{GpRasWil}. \\
In this letter, we present Gaussian processes for various physics applications to highlight their importance and usefulness.
Compared to many other Machine Learning methods, Gaussian processes provide interpretable, probabilistic results with quantified uncertainty while being related to many commonly used models such as neural networks and splines \cite{nn_gp, spline_gp}.
We improve systematic corrections of an electron energy detector by modeling underlying phenomena with Gaussian processes and reconstructing the electron signal.
Furthermore, we use Gaussian processes in sequential optimization for further fine-tuning detector parameters.
We achieve optimal detector performance with both corrections while highly reproducible and using the fewest samples possible.
We recommend advanced Gaussian process models such as stochastic variational Gaussian processes and discuss example applications for optimization and modeling in particle physics experiments.
\newline
\\
For our primary example, consider a simplified electron energy detector for an incoming electron with kinetic energy $E_e$.
The detector is made up of a square scintillator, light guides at the sides, and a photomultiplier tube (PMT) at the end of each light guide, as shown in Fig.~\ref{fig:det_schematic}.
When an electron hits the scintillator, it deposits its energy, and $n_{\gamma}$ photons are produced in the scintillator proportional to the electron energy $E_e$.
Photons travel through the detector to the PMTs and are converted to charge pulses with total charge $A$.
The measured charge $A$ is on an arbitrary scale, and we assume a linear relation to the original electron energy
\begin{align*}
    & A \propto n_{\gamma} \propto E_e \\
    & E_e = g \cdot A
\end{align*}
with gain $g$.
We calibrate the gain with a mono-energetic electron source with known energy. \\
We must expand our simple model for an actual data set with two systematic effects that need to be corrected.
In our example, we treat each systematic effect independently.
Firstly, the gain $g$ will fluctuate over time due to temperature-induced gain fluctuations in the signal chain
\begin{align}
    g (t) = g \cdot c_T (t) .
    \label{equ:driftcorr}
\end{align}
There is no reliable, analytical model for the temperature dependence of $g (T)$, as PMTs are inherently highly non-linear.
Secondly, the PMTs have not the same gain $g_i$ and need to be fine-tuned with a relative gain factor $c_i$ for each PMT $i$ to get the correct total charge
\begin{align}
    A = \sum_i c_i \cdot A_i .
    \label{equ:spatcorr}
\end{align}
We write all PMT fine-tuning factors as a vector $\mathbf{c}$. PMTs are hard to tune to have the same gain for previously stated reasons and because photon paths in the detector depend on the electron position where it hit the detector.

We can also characterize both effects with a mono-energetic source.
For the temperature-induced gain fluctuations, we measure the total charge $A$ with the source in front of the detector at different times.
For fine-tuning, we measure the electron energy on a grid of different positions in front of the detector.
Such measurements contain more information about individual PMT contributions.
We expect symmetric and uniform results for an ideal detector.
We calculate the measured deviation from symmetry and uniformity from the resulting grid measurement as one scalar $\mathcal{L}$.
To correct a measurement with the described detector, we need a model to infer $c_T (t)$ and $\mathcal{L}(\mathbf{c})$ from data. \\
The data set we use for our example problem stems from a recent measurement with the \perkeo instrument 2019/20 at the Institute Laue-Langevin, Grenoble, France \cite{BastianPerkeoIII, BastiReview}.
In the experiment, cold neutrons decay in a magnetic field, and the emitted electrons are detected with two detectors similar to our schematic.
By measuring the electron energy spectrum precisely, new limits on tensor interactions beyond the Standard Model can be set \cite{Gonz1, Gonz2, UCNA_spec, UCNA_Ab, HeikoFierz, markisch2019measurement}.
\noindent 

We propose using Gaussian processes regression to model such underlying physical corrections of measurements and similar problems like modeling physical processes from a set of simulation data.
We can model the predictions and local uncertainties of a continuous input space by treating each point in it as a component of an infinite-dimensional Gaussian distribution. 
A sample from this Gaussian distribution will be a function with a value for each point in the input space. 
Whereas, each point $\mathbf{x}$ has a mean $m(\mathbf{x})$ and covariance $k(\mathbf{x}, \mathbf{x^\prime})$ to other points $\mathbf{x^\prime}$.
This defines a Gaussian process with function samples $\mathbf{f}$
\begin{align}
    \mathbf{f} \sim \mathcal{GP} \left( m( \mathbf{x} ), k( \mathbf{x}, \mathbf{x^\prime}) \right) .
\end{align}
The mean and fluctuation of the sample function $\mathbf{f}$ state the predictive mean and uncertainty at a point $\mathbf{x}$. We want to construct such a model using a finite data set of observations. Using Bayesian inference, we can define a prior over functions with zero mean $m(\mathbf{x}) =  0$, covariance matrix $\mathbf{K}$ with entries $K_{ij} = k(\mathbf{x_i}, \mathbf{x_j'})$, an observed data set with input points $X \in \mathcal{R}^{m}$ and values $y \in \mathcal{R}$
\begin{align*}
    \mathbf{f} \sim \mathcal{N}\left(\mathbf{0}, ~\mathbf{K}(X, X) \right)
\end{align*}
We can expand this for observations with fixed noise $y = f(\mathbf{x}) + \epsilon$ where $\epsilon \sim \mathcal{N}(0, \sigma_n^2)$. In addition, we add testing points $X_{*}$ in the input space where we would like to predict function values $\mathbf{f}_{*}$ to get the joint prior over functions
\begin{align}
    \left[\begin{array}{l}
    \mathbf{y} \\
    \mathbf{f}_{*}
    \end{array}\right] \sim \mathcal{N}\left(\mathbf{0},\left[\begin{array}{cc}
    \mathbf{K}(X, X)+\sigma_{n}^{2} \mathbf{I} & \mathbf{K}\left(X, X_{*}\right) \\
    \mathbf{K}\left(X_{*}, X\right) & \mathbf{K}\left(X_{*}, X_{*}\right)
    \end{array}\right]\right) .
    \label{equ:jointprior}
\end{align}
As we are using Gaussian distributions, we do not need to calculate the posterior distribution through Bayes rule but can obtain the predictive distribution through conditioning
\begin{align}
    \mathbf{f}_{*} \mid X, \mathbf{y}, X_{*} \sim \mathcal{N}\left(\overline{\mathbf{f}}_{*}, \operatorname{cov}\left(\mathbf{f}_{*}\right)\right) ,
\end{align}
with predicted mean $\overline{\mathbf{f}}_{*}$ and covariance $\operatorname{cov}\left(\mathbf{f}_{*}\right)$ at testing points $X_{*}$
\begin{equation}
    \begin{aligned}
        \overline{\mathbf{f}}_{*} &= \mathbf{K}\left(X_{*}, X\right)\left[\mathbf{K}(X, X) + \sigma_{n}^{2} \mathbf{I}\right]^{-1} \mathbf{y} \\
        \operatorname{cov}\left(\mathbf{f}_{*}\right) &= \mathbf{K}\left(X_{*}, X_{*}\right) \\
            &\quad\quad - \mathbf{K}\left(X_{*}, X\right)\left[\mathbf{K}(X, X)+\sigma_{n}^{2} \mathbf{I}\right]^{-1} \mathbf{K}\left(X, X_{*}\right) .
        \label{equ:predmeancov}
    \end{aligned}
\end{equation}
Thus, Gaussian processes reconstruct the underlying signal without contaminating noise by computing a weighted average of noisy observations $\mathbf{y}$. 
Therefore, Gaussian processes are equivalent to linear smoother. Furthermore, they are also a generalization of spline models, and large neural networks as well as support vector machines\cite{nn_gp, spline_gp, GpRasWil}. 
They can be applied to regression (GPR) or classification tasks. \\
The only parameters of a Gaussian process described above are the noise parameter $\sigma_n$ and any parameters of the covariance function $k$. Therefore, Gaussian processes are non-parametric, as they are created from data sets themselves and grow in complexity with them.
We can set these parameters by maximizing the marginal log likelihood $\log p(\mathbf{y} \mid X)$ obtained through the Gaussian likelihood and prior as
\begin{equation}
    \begin{aligned}
        \log p(\mathbf{y} \mid X) &= -\frac{1}{2} \mathbf{y}^{\top}\left(\mathbf{K}(X, X)+\sigma_{n}^{2} \mathbf{I}\right)^{-1} \mathbf{y} \\
        &\quad\quad\quad - \frac{1}{2} \log \left|\mathbf{K}(X, X) + \sigma_{n}^{2} \mathbf{I}\right|-\frac{n}{2} \log 2 \pi .
    \end{aligned}
    \label{equ:mll}
\end{equation}
The noise and kernel function parameters are optimized during the fit, implying interpretable results, unlike other machine learning approaches. \\

\begin{figure}[t]
    \includegraphics[width=0.4\textwidth]{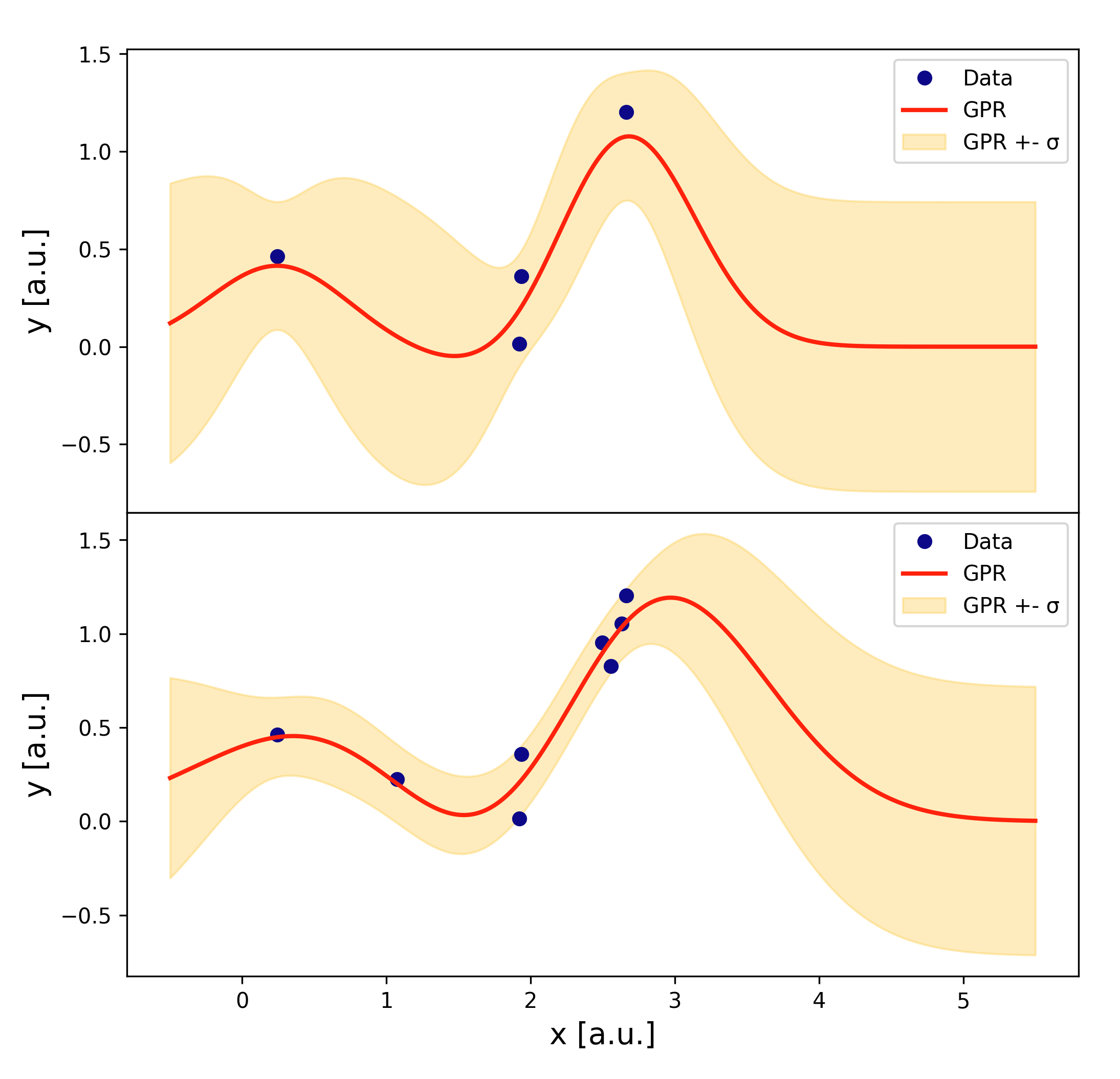}
    \caption{\label{fig:gpr_demo} Example: Gaussian process posterior distribution for two different data sets. One with 4 (top) and updated for 8 (bottom) noisy points of the same unknown underlying distribution.}
\end{figure}

To illustrate the non-parametric nature, we give an example GPR for a one-dimensional input space for a given data set with four and eight data points in Fig.~\ref{fig:gpr_demo}.
Note that the predicted mean and covariance reflect the prior again away from the data. 
The predictive range depends on the length scale parameter of the kernel function, see Equ.~\eqref{equ:rbf}. \\

The covariance function is the most essential model assumption when using Gaussian processes. 
Kernel functions $k$ are integral operators and are valid covariance functions, if positive semi-definite \cite{kernel_scholkopf2018}. 
The matrix $\mathbf{K}$ with entries $K_{ij} = k(\mathbf{x_i}, \mathbf{x_j'})$ is then a covariance matrix.
Kernel functions score points in the input space according to their similarity with a given measure. 
This scoring can be interpreted as a projection of $n$ data points in $\mathcal{R}^m$ into a \textit{similarity} space or \textit{kernel} space where every data point is scored to each other point
\begin{align}
    k: \mathbb{R}^{n \times m} \rightarrow \mathbb{R}^{n \times n} .
    \label{equ:kerneldim}
\end{align}
Mathematically, this is also equivalent to projecting $\mathbf{x}$ and $\mathbf{x^\prime}$ into a usually higher dimensional feature space with measure $\nu$ via a mapping function $\phi$ and taking the inner product 
\begin{align}
    k(\mathbf{x}, \mathbf{x^\prime}) = \langle \phi(\mathbf{x}), \phi(\mathbf{x^\prime}) \rangle_{\nu} .
\end{align}
However, neither the projection nor the inner product needs to be calculated directly \cite{GpRasWil, kernel_scholkopf2018}. 
This equivalence is referred to as \textit{kernel trick}. 
Kernel functions can be stationary $k(\mathbf{x}, \mathbf{x^\prime}) = k ( \mathbf{x} - \mathbf{x^\prime} )$, i.e. invariant under translations in the input space, or even isotropic $k(\mathbf{x}, \mathbf{x^\prime}) = k ( \left| \mathbf{x} - \mathbf{x^\prime} \right|)$.
Composite kernel functions are also possible: A sum or a product of two kernel functions is also a kernel function.
This characteristic can be seen as logical operators for similarity criteria.
Combining a stationary and a non-stationary kernel will create a composite non-stationary kernel. 
We can also encode symmetry into kernel functions, e.g. as $k(\mathbf{x}, \mathbf{x^\prime}) = k(\mathbf{x}, \mathbf{x^\prime}) + k(-\mathbf{x}, \mathbf{x^\prime})$.
All of these characteristics enable physically motivated kernel functions for different problems. \\
Commonly used kernel functions for Gaussian processes \cite{GpRasWil, BayOpt2} are the radial basis function kernel
\begin{align}
    k(\mathbf{x}, \mathbf{x^\prime}) = \sigma^2 \exp \left( -\frac{ \left|\left| \mathbf{x} - \mathbf{x^\prime} \right|\right|^2} {2 \rho^2}\right)
    \label{equ:rbf}
\end{align}
and the \textit{Matérn} kernel with $d = \left|\left| \mathbf{x} - \mathbf{x^\prime} \right|\right|^2$ 
\begin{equation}
    \begin{aligned}
        k(\mathbf{x}, \mathbf{x^\prime})_{\nu=\frac{3}{2}} &= \sigma^2 \left(  1 + \frac{\sqrt{3}d}{\rho} \right) \exp{\left(  - \frac{\sqrt{3}d}{\rho} \right)} \\
        k(\mathbf{x}, \mathbf{x^\prime})_{\nu=\frac{5}{2}} &= \sigma^2 \left(  1 + \frac{\sqrt{5}d}{\rho} + \frac{5 d^2}{2 \rho^2}\right) \exp{\left(  - \frac{\sqrt{5}d}{\rho} \right)}.
    \end{aligned}
    \label{equ:matern}
\end{equation}
Both have a length scale parameter $\rho$, an output scale parameter $\sigma$ and are isotropic kernel functions. 
Furthermore, the differentiability of a kernel function affects the smoothness of the samples drawn from the corresponding Gaussian process prior in Equ.~\eqref{equ:jointprior}.
Naturally occurring phenomena usually have underlying non-continuous contributions.
Smooth samples can struggle to model such phenomena, leading to a length scale collapse when maximizing the marginal log likelihood. 
This collapse, in turn, yields imprecise posterior distributions.
The RBF kernel function in Equ.~\eqref{equ:rbf} is infinitely mean-square differentiable, whereas the Matérn kernel function in Equ.~\eqref{equ:matern} is only $ \left\lceil \nu \right\rceil - 1$ times differentiable \cite{kernel_scholkopf2018}. \\

\begin{figure}[t]
    \includegraphics[width=0.5\textwidth]{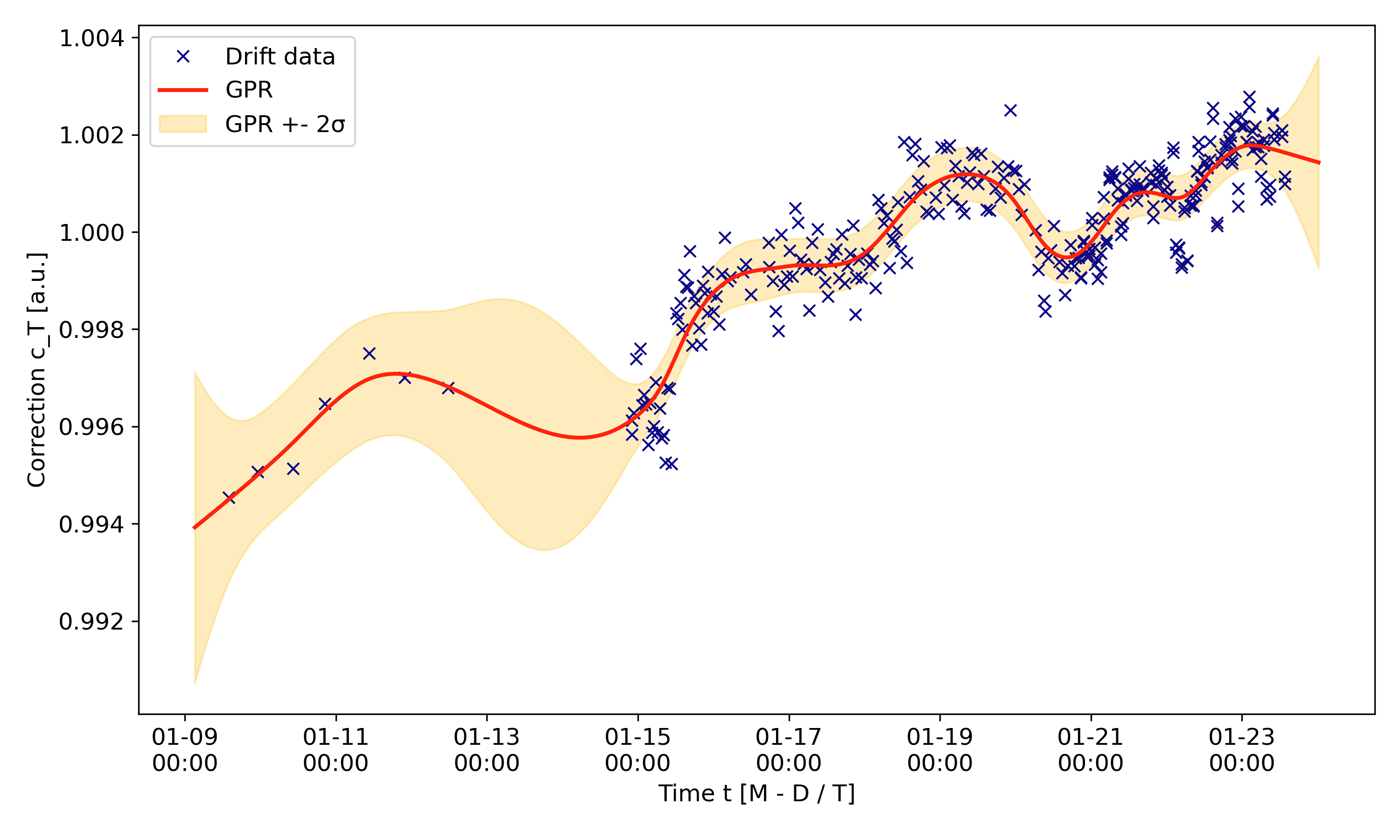}
    \caption{\label{fig:gpr_drift} Temperature-induced gain fluctuation correction $c_T (t)$ with Gaussian process regression, reconstructing the underlying signal of the fluctuations.}
\end{figure}

With the Gaussian process model, we can now reconstruct underlying signals in a non-parametric and fully Bayesian way.
Hence, we apply GPR with a Matérn kernel function ($\nu = 3/2$) from Equ.~\eqref{equ:matern} and a mean prior $m(\mathbf{t}) = 1$ to determine the temperature induced gain fluctuation correction $c_T (t)$ at different times $t$ from Equ.~\eqref{equ:driftcorr}.
We use hourly measurements with a mono-energetic source of known energy between actual measurements with the detector.
We assume a constant, actual gain, which we extract as the weighted arithmetic mean of the hourly measurements.
The relative deviation from the weighted arithmetic mean is the pointwise correction ${c_T}^{\prime}(t_i)$ for the time of measurement $t_i$.
Therefore, the actual correction is given by 
\begin{align*}
    c_T(t_i) = {c_T}^{\prime}(t_i) + \epsilon \quad\text{with}\quad \epsilon \sim \mathcal{N}(0, \sigma_n^2)
\end{align*}
in full analogy to Equ.~\eqref{equ:jointprior}.
We use GPR to avoid correcting for statistical fluctuations of the measurements.
The used data set and the GPR results are shown in Fig.~\ref{fig:gpr_drift}.
The GPR model reconstructs the underlying correction and gives its statistical uncertainty.
The predicted uncertainty enables accurate quantification of the systematic error on the final measurement.
We use this to optimize the time window selection of the data set used for the final measurement to achieve higher precision of experimental results.
Furthermore, the GPR model parameters are interpretable, as the noise parameter $\sigma_n$ states the statistical fluctuation of the pointwise corrections ${c_T}^{\prime}$ and the kernel function length scale $\rho$ states the timescale for correction changes to occur. 
However, we set the validity of the interpretability with the choice of the covariance function.
A smooth kernel does not accurately model underlying non-continuities, and a stationary kernel fails to model data with local length scales $\rho$, even if we set the parameters by optimizing the marginal log likelihood in Equ.~\eqref{equ:mll}. \\

\begin{figure}[t]
    \includegraphics[width=0.45\textwidth]{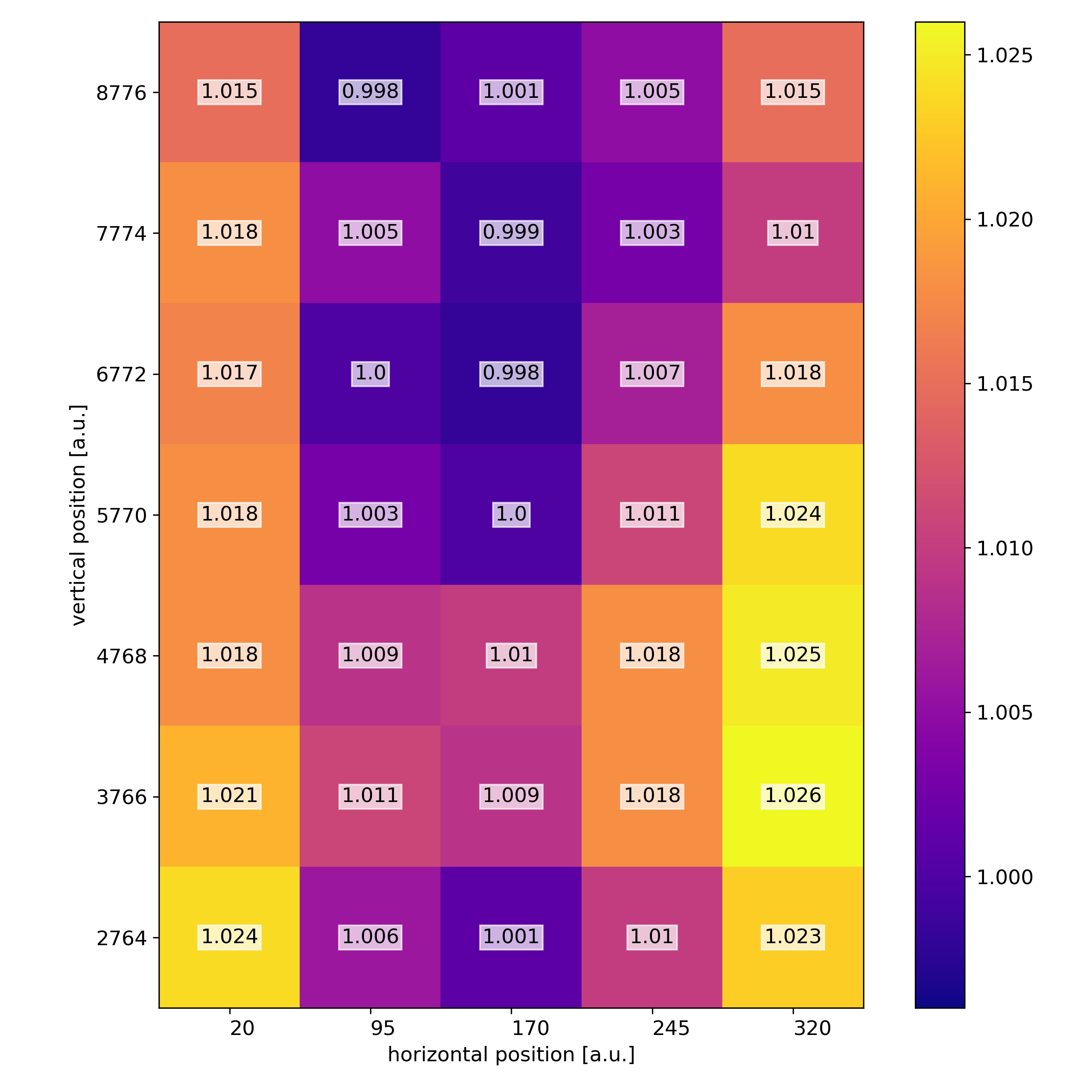}
    \caption{\label{fig:map_unopt} Unoptimized spatial response map with the objective values from Equ.~\eqref{equ:totloss} with uniformity term $\mathcal{L}_{\text{u}} = 22690$ and symmetry term $\mathcal{L}_{\text{s}} = 9250$. All peak positions $\mu_{i}$ are shown relative to $\mu_{\text{c}}$.}
\end{figure}

For our second correction, we want to determine ideal fine-tuning or gain factors $\mathbf{c}$ for each PMT $i$ out of all eight to get the correct total charge $A$ in Equ.~\eqref{equ:spatcorr}.
For fine-tuning, we use a grid of measurements for different positions in front of the detector, with a mono-energetic source of known energy.
At each position $k$, we measure a spectrum of total charges $A^{\prime} = A (\mathbf{c})$ for several electrons and extract the peak position $\mu_k$ with a fit of the histogram.
For an ideal detector, the results for the grid points are symmetric and uniform.
An ideal fine-tuning should also lead to a uniform and symmetric result. 
We could calibrate the detector without fine-tuning $\mathbf{c} = \mathbf{1}$. However, this would mean accepting non-uniform and asymmetric results.
Such results would need to be determined accurately, as we must correct other effects depending on detector positioning or the incoming electron distribution in reality.
Usually, the required accuracy for this is problematic. \\
We illustrate the grid of measurements or \textit{spatial response} of the detector as a map in Fig.~\ref{fig:map_unopt} without any fine-tuning. \\

We calculate a measured deviation from symmetry and uniformity from the grid measurement in one scalar quantity $\mathcal{L}$.
We use this quantity as objective function $\mathcal{L} (\mathbf{c})$ to turn it into an optimization problem
\begin{align}
    \mathcal{L} \left( \mathbf{c} \right) = \mathcal{L}_{\text{u}} \left( \mathbf{c} \right) + \mathcal{L}_{\text{s}} \left( \mathbf{c} \right) 
    \label{equ:totloss}
\end{align} 
Minimizing $\mathcal{L}$ yields ideal gain factors $\mathbf{c}$.
The uniformity term $\mathcal{L}_{\text{u}}$ is given by the mean square error of each grid point $k$ for all points $n$ in respect to the peak position $\mu_{\text {c}}$ of the middle point for unoptimized gain factors, i.e. $\mathbf{c} = \mathbf{1}$.
\begin{align}
    \mathcal{L}_{\text{u}}  \left( \mathbf{c} \right) =\frac{1}{n} \sum_{k}\left(\mu_{\text{c}}-\mu_{k}\left( \mathbf{c} \right) \right)^{2}
    \label{equ:uloss}
\end{align}
The reference point $\mu_{\text {c}}$ maintains a similar energy-channel relation before and after the optimization by avoiding significant changes in the overall detector gain.
To also favor symmetric solutions, the square deviations from the $m$ points in the left and rightmost columns left to right and top to bottom are added as symmetry loss $\mathcal{L}_{\text{s}}$.
The symmetric deviation is only calculated for these columns, as they are closest to the PMTs and therefore most sensitive to any individual gain changes
\begin{equation}
    \begin{aligned}
        \mathcal{L}_{\text{s}}  \left( \mathbf{c} \right) &= \alpha_1 \sum_{k}\left(\mu_{\text {left}, k}\left( \mathbf{c} \right)-\mu_{\text{right}, k}\left( \mathbf{c} \right)\right)^{2} \\
        &\quad\quad\quad +  \alpha_2 \sum_{k}\left(\mu_{\text{top}, k}\left( \mathbf{c} \right)-\mu_{\text{bot}, k}\left(\mathbf{c} \right)\right)^{2} ,
    \end{aligned}
    \label{equ:sloss}
\end{equation}
with scaling factors $\alpha_1 = 0.5 / m$ and $\alpha_2 = 0.5 / (m - 1)$.
For a given set of gain factors $\mathbf{c} \in \mathbb{R}^8$ and the corresponding objective function value $\mathcal{L}$, we need to recalculate all spectra for all grid points with signal processing and consecutive fit.
This process requires some time, and any solution method would need to guarantee symmetric and reproducible results.
In addition, the \perkeo data set has many of these independent spatial response measurements to optimize, emphasizing the use of as few objective function calls as possible, i.e., high sample efficiency.
However, the parameters $\mathbf{c}$ highly correlate, and the problem is hard to solve without brute force.
To simplify, we can combine the PMTs and their gain factors $c_i$ in symmetric groups of two (each corner).
This grouping reduces dimensionality and correlations of the problem by creating a fixed, embedded subspace of the original $\mathbb{R}^8$ input space at the cost of minima quality.
We will use this later to further compare different methods and their performances. \\

To predict minima of the objective function $\mathcal{L} (\mathbf{c})$ in Equ.~\eqref{equ:totloss}, we require a model to infer values of $\mathcal{L}$ for untested gain factors $\mathbf{c}$. 
With an accurate model of $\mathcal{L} (\mathbf{c})$, we can evaluate it with little cost, needing fewer tested samples of $\left( \mathbf{c}, ~\mathcal{L} \right)$ pairs. 
Selecting the next test point $n + 1$ by leveraging the uncertainty of the current model for a data set $\mathcal{D}_n = \{  (\mathbf{c}_l, \mathcal{L}_l) \mid l = 1, ...., n \}$ of $n$ tested points sequentially is called \textit{Bayesian optimization} \cite{BayOpt2}.
This sequential approach yields a fully Bayesian model-based data-driven method to optimize any problem requiring high sample efficiency, being it due to time or resource constraints. \\

Bayesian optimization with GPR works without knowing the to-be-optimized black-box function analytically, removing the expensive amount of function evaluations gradient-based approaches and other methods require.
This method found applications in hyperparameter tuning \cite{bergstra2011algorithms}, outperforming other methods, and has also found physics applications \cite{bayoptslacacc, bayoptslac2, bayoptdesi, bayoptcosmo, bayopthiggs, bayopteventgen, bayoptdetectordesign, bayoptnuclear}.
Furthermore, the authors of \cite{SlacMobo} recently demonstrated how the approach could be expanded to multi-objective Bayesian optimization (MOBO) \cite{MoboBook} to optimally balance the trade-offs between multiple competing objectives for physics applications simultaneously.

To determine the next set of untested gain factors $\mathbf{c}$ to be sampled, we can formulate an \textit{acquisition function} that scores candidate data points' utility for the next evaluation, guiding the sequential search. 
Acquisition functions need to balance exploration of the input space versus exploitation of the current best result to achieve overall sample efficiency and also reach the global optimum.

For our systematic correction, we use GPR with a Matérn kernel function ($\nu = 5/2$) from Equ.~\eqref{equ:matern}.
We employ a combination of improvement-based and optimistic policies as acquisition function by mainly using expected improvement (EI) and lower confidence bounds (LCB) \cite{BayOpt2}.
When the EI utility is below a threshold and could lead to stagnation, we use LCB.
EI is generally more often used as acquisition function \cite{BayOpt2}. 
For GPR posterior distribution with standard deviation $\sigma_n$ and mean prediction $\mu_n$ for given dataset $\mathcal{D}_n$, both acquisition functions are tractable. The LCB is
\begin{align}
    \alpha_{\text{LCB}} (x ; \mathcal{D}_n) &= \mu_n(x) - \kappa \cdot \sigma_n(x)
\end{align}
with the exploration parameter $\kappa$. We set $\kappa = 3$ for more exploration. For GPR posterior distribution $f$ with normal distribution $\mathcal{N}$ and its cumulative distribution function $\Psi$, EI is
\begin{equation}
    \begin{aligned}
        \alpha_{\text{EI, GP}} (x ; \mathcal{D}_n)
        &= \mathbb{E}_{} \max{\left( f(x) - f(x^+_n), 0 \right)} \\
        &= \sigma_n(x) \cdot \gamma(x) \cdot \Psi \left( \gamma(x) \right) + \mathcal{N} \left( \gamma(x) \right)
    \end{aligned}
\end{equation}
with $\gamma(x) = \left(f(x) - f(x^+_n) - \kappa \right) / \sigma_n(x)$, exploration parameter $\kappa$.
We found $\kappa = 1$ to balance a greedy search with the necessary exploration for the problem. The authors of \cite{BayOpt2, e3i} present other possible acquisition functions, like entropy-based acquisition functions.
Finally, we determine the best next candidate to be sampled $x_{n + 1}$ with a gradient-based approach on the acquisition function.
We seed the approach with the best current optimum and random points. \\

To compare Bayesian optimization with GPR, we also use empirically tuned random normal sampling ($\mathbf{c} \sim \mathcal{N}(\boldsymbol{\mu} = \boldsymbol{1}, \boldsymbol{\Sigma} = 0.05 ~ \mathbf{I})$) as baseline, a simulated annealing algorithm and a tree-structured Parzen estimator (TPE) \cite{bergstra2011algorithms}.
TPE is also a case of Bayesian optimization, but instead of modelling the posterior distribution $p \left( y | x \right)$, it models the likelihood $p \left( x | y \right)$.
The TPE model is one of the standard sequential model-based optimization approaches for machine learning hyperparameter optimization, as it is less computationally heavy to compute than, e.g., GPR \cite{bergstra2011algorithms}.
However, due to the covariance function, GPR is naturally more able to deal with highly correlated problems than TPE.
As an alternative approach to modeling the posterior distribution, we tried the novel deep adaptive design (DAD) method \cite{dad}, a Bayesian optimal experimental design \cite{boed_lindley} approach where an ideal optimization sequence for a given number of steps is learned, enabling rapid online inference and eliminating costly posterior update calculations.
DAD would be useful for our example to have an optimization policy for all maps. However, the training turned out to be too costly.
All code for this publication is available on \href{https://github.com/maxlampe/detector-bayopt}{\textit{GitHub}}\footnote{\url{https://github.com/maxlampe/detector-bayopt}}.
The spatial response maps are calculated with the open \perkeo analysis package \href{https://github.com/maxlampe/panter}{\textit{panter}}.
The Bayesian optimization algorithm is implemented with the probabilistic programming language \textit{Pyro} \cite{bingham2018pyro}, which builds on \textit{GPyTorch} \cite{gardner2018gpytorch} and we used the already implemented TPE from the \textit{Optuna} \cite{optuna_2019} package. \\

\begin{figure}[t]
    \includegraphics[width=0.45\textwidth]{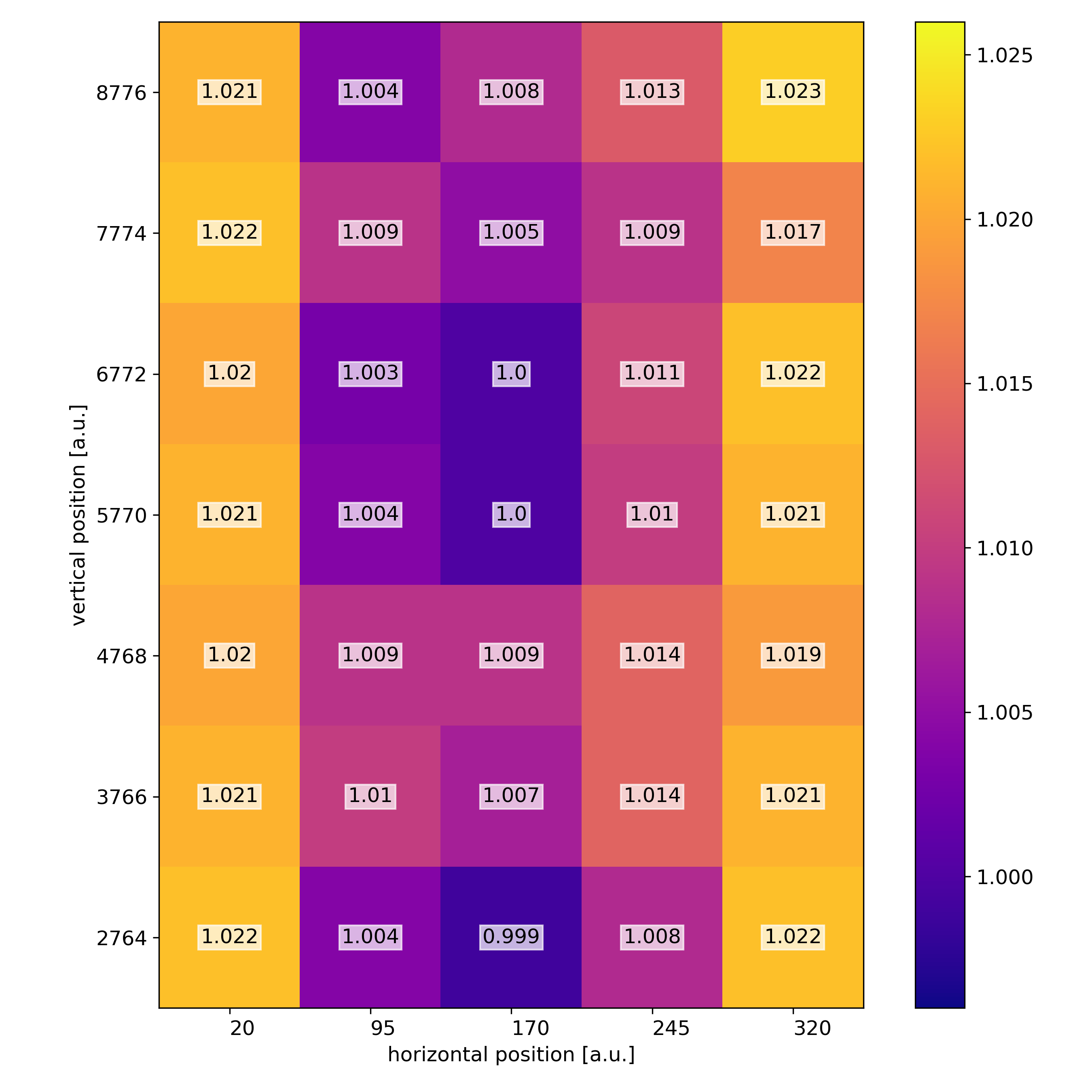}
    \caption{\label{fig:map_opt} Optimized spatial response map with GPR of Fig.~\ref{fig:map_unopt} with the objective values from Equ.~\eqref{equ:totloss} with uniformity term $\mathcal{L}_{\text{u}} = 6360$ and symmetry term $\mathcal{L}_{\text{s}} = 710$. All peak positions $\mu_{i}$ are shown relative to $\mu_{\text {c}}$. Same color scaling as Fig.~\ref{fig:map_unopt}.}
\end{figure}

Symmetric uniformity is the overall goal, and for this problem, a minimized $\mathcal{L}_{\text{s}}$ implies a small $\mathcal{L}_{\text{u}}$, but not vice versa.
Therefore, we use total loss $\mathcal{L}$ from Equ.~\eqref{equ:totloss} as objective value during the optimization and symmetry loss $\mathcal{L}_{\text{s}}$ for comparing results.
Besides symmetry loss, we chose function evaluations as an additional metric.
One function evaluation means calculating one pair $\left( \mathbf{c}, ~\mathcal{L} (\mathbf{c}) \right)$.
We initialize the GPR with a few random data points from the random normal sampler, which we count as function evaluations.
Using function evaluations as metric renders simulated annealing useless, as it requires an order of magnitude more.
As mentioned before, we can group the individual PMT gain factors to compare different degrees of complexity.
This grouping means that we can exploit the effective dimensionality of the problem at the cost of the overall optimization result.
Tab.~\ref{tab:comp_res} presents the results for different PMT groupings and fixed function evaluations.
One optimized spatial response map is shown in Fig.~\ref{fig:map_opt}. \\

\begin{table*}
\caption{\label{tab:comp_res} Comparing optimization results for different methods and PMT groupings, i.e., different problem dimensionality. We fixed the number of function evaluations and run five uncorrelated optimizations for each setting for the map in Fig.~\ref{fig:map_unopt}, with the best achieved value $\mathcal{L}_{\text{s}}^*$ and the average achieved value $\bar{\mathcal{L}}_{\text{s}}$.}
\begin{ruledtabular}
\begin{tabular}{lclccc}
\textrm{Dimensionality}&
\textrm{Function evaluations}&
\textrm{Method}&
\textrm{$\mathcal{L}_{\text{s}}^*$}&
\textrm{$\bar{\mathcal{L}}_{\text{s}}$}&
\textrm{$\bar{\mathcal{L}}_{\text{s}} - \mathcal{L}_{\text{s}}^*$}\\
\colrule
 & & Unoptimized & 9251\\
\colrule
$\mathbf{c} \in \mathbb{R}^4$ & 75 & Random normal & 1070 & 2900 & 1830 \\
    & & TPE  & 1200 & 3800 & 2600 \\
    & & $\textbf{GPR}$ & $\boldsymbol{980}$ & $\boldsymbol{995}$ & $\boldsymbol{14}$ \\
\colrule
$\mathbf{c} \in \mathbb{R}^8$ & 150 & Random normal &  4130 & 4640 & 510 \\
    & & TPE & 1540 & 2590 & 1050 \\
    & & $\textbf{GPR}$ & $\boldsymbol{643}$ & $\boldsymbol{675}$ & $\boldsymbol{32}$ \\
\end{tabular}
\end{ruledtabular}
\end{table*}

\begin{figure}[t]
    \includegraphics[width=0.45\textwidth]{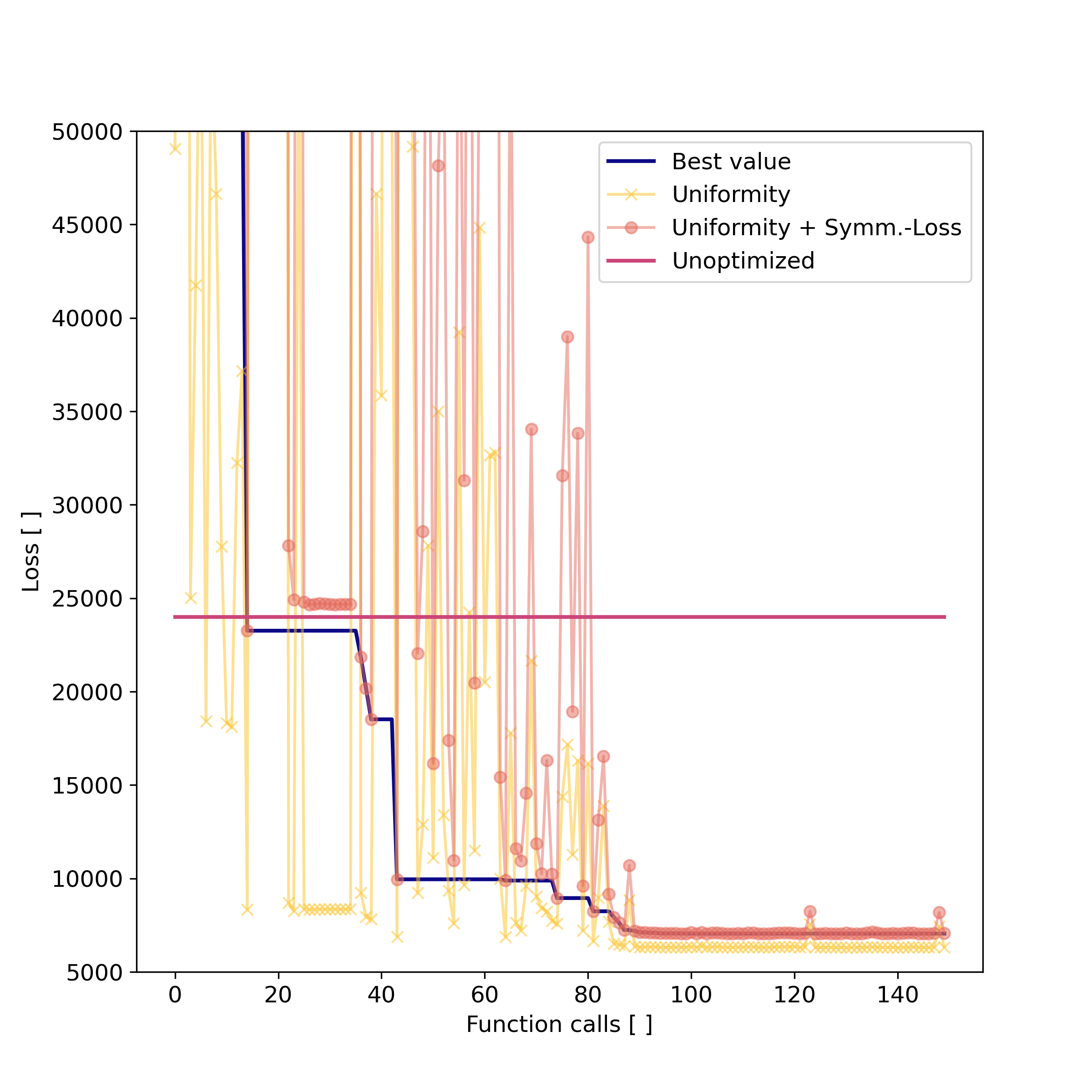}
    \caption{\label{fig:bayopt_loss_curve} Optimization history with GPR for one run of Tab.~\ref{tab:comp_res}. The current best optimization result for the given number of function calls is shown, as well as uniformity and symmetry loss terms.}
\end{figure}

Bayesian optimization with Gaussian processes outperforms the other methods in overall best-achieved results and reproducibility. 
This difference is due to the capability of Gaussian processes to model highly correlated problems accurately.
Contrary, the TPE approach suffers a high spread of results in Tab.~\ref{tab:comp_res}.
It is noteworthy how a well-tuned random normal sampler outperforms the TPE approach, as the modeled likelihood requires a large enough data set for this problem.
For all uncorrelated optimization runs, GPR would have required even fewer function calls as given, highlighting its outstanding sample efficiency.
This effect is shown in the optimization history for one run in Fig.~\ref{fig:bayopt_loss_curve}, which represents all runs.
With this approach, we successfully correct the spatial response of \perkeo detectors to achieve unreached symmetry and uniformity for any scintillation detector previously seen in the collaboration.
These results enable more precise systematic corrections for the overall measurement and, therefore, higher overall precision.
We can interpret the length scale parameter of the covariance function as inherent step size for the optimization problem. \\

%
%

Unfortunately, Gaussian processes suffer from computational complexity of $\mathcal{O}(n^3)$ for a dataset of size $n$, due to the matrix inversion in the marginal log likelihood in Equ.~\eqref{equ:mll}. 
The scaling issue motivated different approximation methods, and we want to highlight what we believe to be the most promising for physics applications: Stochastic variational Gaussian processes (SVGP) \cite{svgp_hensman2013, svgpclass_hensman2015}.
SVGP adds inducing points $X_{\text{S}}$ as parameters to the model, whereas the number of inducing points is smaller than the number of data points $X$.
We optimize the model with a lower bound on the marginal likelihood to fit it and have $(X_{\text{S}}, ~\mathbf{f}_{\text{S}} (X_{\text{S}}))$ summarize the data set.
We outline the mathematical background of the SVGP model and the usage of variational inference \cite{blei2017variational} in the appendix.
With SVGP, we can use Gaussian processes' benefits for large data sets with millions of points.
These benefits are essential for applications where large data sets are necessary to capture the underlying physical phenomena accurately, like simulating experiments or many particle physics applications in general.
A surrogate model from simulation data can be used to get an integrable model that accurately models the underlying phenomena.
Therefore, we can also model individual parts of an experiment when many simulation parameters are present, and a complete mapping of the parameter space would be too costly. 
We successfully apply SVGP to infer from fixed simulation points of the \perkeo detector response to the entire parameter space. 
We can accurately derive energy-dependent systematic corrections with uncertainty to refine the \perkeo results. 
Other useful expansions are Gaussian process latent variable models (GP-LVM) \cite{gplvm_lawrence2003, gplvm}, deep kernel learning with Gaussian processes (DKL-GP)\cite{dkl_wilson2016} and deep Gaussian processes (DGP) \cite{deepGP}. 
GP-LVM create a lower-dimensional non-linear embedding of high dimensional data, whereas DKL-GP use the adpative basis functions of neural networks to enable non-euclidean similarity metrics throughout the input space. DGP stack Gaussian processes, expanding the simple metrics of kernel functions to use different metrics in different input space domains, with a hierarchical feature extraction structure. 
There are also stochastic variational expansions for better scalability of these models available, such as stochastic variational DKL (SV-DKL) \cite{svdkl_wilson2016}.
We only present scalar outputs in this letter, however, multi-output Gaussian processes are also possible \cite{mo_gp}.\\

%
%

Due to the flexibility and inherent probabilistic, non-parametric nature, Gaussian processes remain state-of-the-art for uncertainty quantification tasks \cite{shi2020sparse, ABDAR2021243} and are ideal for experimental physics applications.
We can incorporate underlying physical models with an appropriate choice of kernel functions.
Furthermore, interpretable model parameters set Gaussian processes apart from other machine learning methods.
We believe that Gaussian processes and Bayesian optimization are strongly underused, in contrast to their great potential. 
They enable more time to spend on actual physics instead of analyzing systematic corrections.
Furthermore, Gaussian processes provide well-calibrated predictive uncertainty for many tasks and are well-suited as an input model for Bayesian optimization.
This letter aims to highlight the advantages of the methods and establish them as tools for experimental data analysis.
Using GPR, we improve the \perkeo analysis by modeling systematic corrections directly or optimizing analysis parameters.
We outperform previous analyses in terms of systematic uncertainty quantification and achieve the best spatial response correction for the \perkeo collaboration while being sample efficient and reproducible \cite{HeikoDiss, HeikoFierz}.
In doing so, we improve the overall precision and quality of the analysis. \\
Beyond \perkeo, Bayesian optimization with Gaussian processes found applications in optimizing detector design parameters \cite{bayoptdetectordesign}, optimizing and online adjust experimental setups, so far only used for accelerators \cite{bayoptslacacc, bayoptslac2, SlacMobo, bayoptdesi}, and for setting analysis and simulation parameters \cite{bayopteventgen, bayoptcosmo, bayoptnuclear, bayopthiggs, bayoptchem}.
Gaussian processes found applications in astrophysics \cite{astroGP, gp_cosmo, gp_gravwaves, gp_pulsars}, condensed matter physics \cite{SGP_condensed} and dynamical systems \cite{gp_moleculardynamics, gp_neurovascular}, but few in particle physics.
One reason is the size of data sets in particle physics applications. 
Expanded models like SVGP, GP-LVM, DKL and SV-DKL should diminish previous limitations and issues.
We outline promising applications:
Gaussian processes can model systematic corrections directly, e.g., reconstructing unknown background signal time series, inferring energy-dependent corrections from simulations, or modeling spatial dependencies.
Bayesian optimization with GPR can, e.g., optimize complex experiment setups and tune its parameters to increase the signal-to-noise ratio. To maximize energy resolution, we can determine thresholds and energy cuts in correlated, costly analysis.
We can use Gaussian processes for classification and apply them to event reconstruction, even to large data sets with SVGP or large data sets with non-euclidean metrics with SV-DKL.  \\
%
\begin{acknowledgements}
    Funded by the Deutsche Forschungsgemeinschaft (DFG, German Research Foundation) under Germany´s Excellence Strategy – EXC 2094 – 390783311.
    We thank D. Greenwald, TUM, for valuable discussions.
\end{acknowledgements}

\section*{Appendix}

We want to outline how stochastic variational Gaussian processes (SVGP) are derived differently from \textit{exact} Gaussian processes, as described in Equ.~\eqref{equ:predmeancov} and \eqref{equ:mll}.
The goal is to have Gaussian processes and their strong modeling capabilities with uncertainty and interpretability scale to large data set applications. 
As highlighted earlier, Gaussian processes suffer from computational complexity of $\mathcal{O}(n^3)$ for a dataset of size $n$, due to the matrix inversion in the marginal log likelihood in Equ.~\eqref{equ:mll}.
The authors of \cite{sparsegp} use a sparse prior with inducing points $X_{\text{S}}$ as additional parameters to model
\begin{align}
    \left[\begin{array}{l}
    \mathbf{f} \\
    \mathbf{f}_{\text{S}}
    \end{array}\right] \sim \mathcal{N}\left(\mathbf{0},\left[\begin{array}{ll}
    \mathbf{K}(X, X) & \mathbf{K}\left(X, X_{\text{S}}\right) \\
    \mathbf{K}\left(X_{\text{S}}, X\right) & \mathbf{K}\left(X_{\text{S}}, X_{\text{S}}\right)
    \end{array}\right]\right) .
    \label{equ:sparseprior}
\end{align}
The idea is to have the inducing points and their predictions $(X_{\text{S}}, ~ \mathbf{f}_{\text{S}} (X_{\text{S}}))$ summarize the data set. In the exact Gaussian process joint prior over functions in Equ.~\eqref{equ:jointprior}, we want to use the information of $\mathbf{y} (X)$ to infer values $\mathbf{f}_{*} (X_{*})$ during inference. 
With the sparse prior, we want to use the information of $\mathbf{f}_{\text{S}} (X_{\text{S}})$ to infer values $\mathbf{f} (X)$ during optimization.
However, using the sparse prior from Equ.~\eqref{equ:sparseprior} results in a marginal log likelihood $\log p(\mathbf{y} \mid X)$ still containing $\mathbf{K}(X, X)$ terms and also losing terms containing the added inducing points $X_{\text{S}}$.
Therefore, we need another objective value to fit our model. \\

To this end, the authors of \cite{svgp_hensman2013, svgpclass_hensman2015} use variational inference \cite{blei2017variational} to solve this problem.
The idea of variational inference is to approximate the posterior distribution with a variational distribution as 
\begin{align*}
    q \left( \mathbf{f}, \mathbf{f}_{\text{S}}   \right)   \approx  p \left(  \mathbf{f}, \mathbf{f}_{\text{S}} \mid \mathbf{y} \right) .
\end{align*}
We can choose the approximation as factorization
\begin{align*}
    q \left( \mathbf{f}, \mathbf{f}_{\text{S}}   \right)  = q \left( \mathbf{f} \mid \mathbf{f}_{\text{S}}   \right)  q \left( \mathbf{f}_{\text{S}}   \right) ,
\end{align*}
with marginal variational distribution $q \left( \mathbf{f}_{\text{S}} \right)$ as Gaussian distribution with variational parameters $\boldsymbol{\mu}$ and $\boldsymbol{\Sigma}$
\begin{align*}
    q \left( \mathbf{f}_{\text{S}} \right)  = \mathcal{N} ( \boldsymbol{\mu}, \boldsymbol{\Sigma} ) .
\end{align*}
The model has inducing points and variational parameters as additional model parameters.
The number of inducing points is set manually and fixed.
For variational inference, we can use a lower bound on the marginal log likelihood as optimization objective, the \textit{evidence lower bound} or $\text{ELBO}$
\begin{align*}
    \log p(\mathbf{y} \mid X)  \geq \mathbb{E}_{\mathbf{f}, \mathbf{f}_{s} \sim q\left(\mathbf{f}, \mathbf{f}_{s}\right)}\left[\log \left(p(\mathbf{y} \mid \mathbf{f}) \frac{p\left(\mathbf{f}, \mathbf{f}_{\text{S}}\right)}{q\left(\mathbf{f}, \mathbf{f}_{\text{S}}\right)}\right)\right] .
\end{align*}
Which we can rewrite as 
\begin{align}
    \text{ELBO} = \sum_{i=1}^{n} \int \log \left(p\left(y_{i} \mid f_{i}\right)\right) q\left(f_{i}\right) \mathrm{d} f_{i} - \text{KL}\left(q\left(\mathbf{f}_{\text{S}}\right) \| p\left(\mathbf{f}_{\text{S}}\right)\right) ,
    \label{equ:elbo}
\end{align}
with the Kullback–Leibler divergence or \textit{relative entropy} $\text{KL}$ \cite{kullback1951information}.
When using the $\text{ELBO}$ as optimization objective with gradient methods, we can approximate the gradient of the sum in the first term in Equ.~\eqref{equ:elbo} with a random subset of the data set. 
This approximation is called stochastic gradient descent (SGD) and avoids using the entire data set. 
The second term is independent of the data set.
Maximizing the $\text{ELBO}$ improves how well the variational distribution approximates the posterior and how well our model describes the original data set. The predictive distribution for the SVGP analog to the exact Gaussian process in Equ.~\eqref{equ:predmeancov} becomes
\begin{align*}
    \mathbf{f}_{*} \mid X, \mathbf{y}, X_{*}, X_{\text{S}} \sim p\left(\mathbf{f}_{*} \mid \mathbf{f}_{\text{S}}\right) 
    =  \mathcal{N}\left(\overline{\mathbf{f}}_{*}, \operatorname{cov}\left(\mathbf{f}_{*}\right)\right) ,
\end{align*}
with new predicted mean $\overline{\mathbf{f}}_{*}$ and covariance $\operatorname{cov}\left(\mathbf{f}_{*}\right)$ at testing points $X_{*}$ and shorthand writing $\mathbf{K}\left(X_{*}, X_{\text{S}} \right) = \mathbf{K}_{*\text{S}}$ etc.
\begin{figure}[t]
    \includegraphics[width=0.5\textwidth]{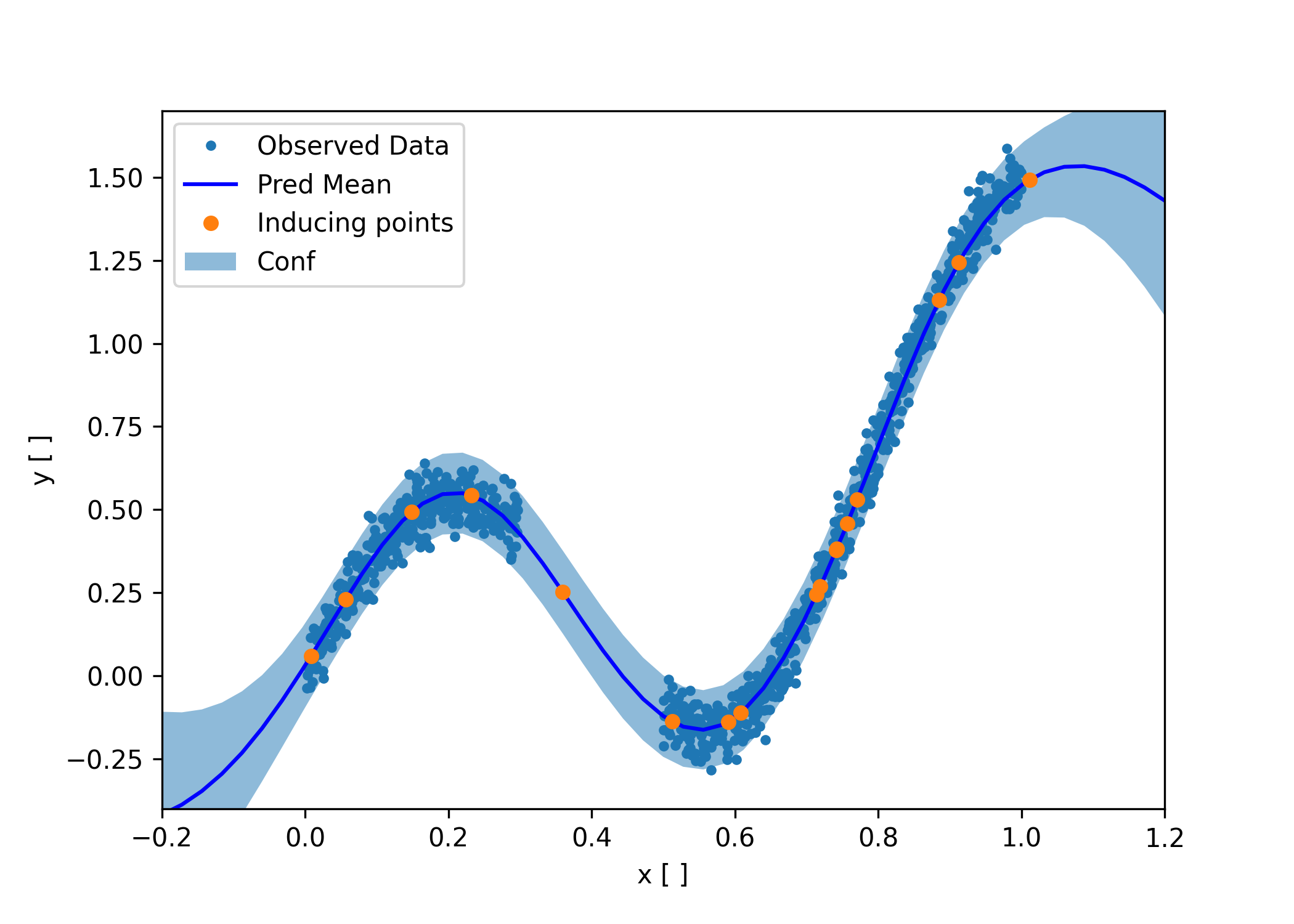}
    \caption{\label{fig:svgp_trained} SVGP example: Predictive distribution with $95\%$ confidence region from Equ.~\eqref{equ:svgppredmeancov} after optimizing the \textit{ELBO} from Equ.~\eqref{equ:elbo}. The 20 inducing points $X_{\text{S}}$ and their predictions $\mathbf{f}_{\text{S}} (X_{\text{S}})$ summarize the data set. We plot only $10\%$ of the observations for illustration.}
\end{figure}
\begin{equation}
    \begin{aligned}
        \overline{\mathbf{f}}_{*} &= \mathbf{K}_{*\text{S}} \mathbf{K}_{\text{S}\text{S}}^{-1} \boldsymbol{\mu} \\
        \operatorname{cov}\left(\mathbf{f}_{*}\right) &= 
        \left( \mathbf{K}_{*\text{S}} \mathbf{K}_{\text{S}\text{S}}^{-1} \right) \boldsymbol{\Sigma} \left( \mathbf{K}_{*\text{S}} \mathbf{K}_{\text{S}\text{S}}^{-1} \right)^\intercal \\
            &\quad\quad\quad\quad +  \mathbf{K}_{**} - \mathbf{K}_{*\text{S}} \mathbf{K}_{\text{S}\text{S}}^{-1} \mathbf{K}_{*\text{S}}^\intercal .
    \end{aligned}
    \label{equ:svgppredmeancov}
\end{equation}
The predictive mean and covariance do not need the original data set anymore. 
We illustrate SVGP on a simple, one-dimensional regression problem with a data set of $\mathcal{O}(n^4)$ points using 20 inducing points $X_{\text{S}}$ in Fig.~\ref{fig:svgp_trained}. 
We generate the data set in Fig.~\ref{fig:svgp_trained} with an arbitrary, underlying function with observational noise.
As can be seen, the inducing points successfully summarize the large data set.
The SVGP model offers all advantages of Gaussian processes but scales to millions of data points, enabling regression and classification applications in particle physics.
Should a physics application require sensitivity to local density changes and non-euclidean similarity metrics, we recommend SV-DKL.
Generally, using variational inference with Gaussian processes is called \textit{variational Gaussian processes} \cite{vgp_blei}, and the approach works for non-Gaussian likelihoods. This approximate inference approach is the key to many Gaussian process expansions, e.g., DGP \cite{deepGP}.

\nocite{*}

\bibliography{apssamp}

\end{document}